\documentclass[preprint,showpacs,preprintnumbers,amsmath,amssymb]{revtex4}  
\usepackage{blindtext}
\usepackage{graphicx}
\usepackage{amssymb}
\usepackage{amsfonts}
\usepackage{amsmath}
\usepackage{textcomp}
\usepackage{xcolor}
\usepackage{ulem}
\usepackage{bm}
\usepackage{parskip,float,graphicx}
\usepackage[caption = false]{subfig}
\usepackage[colorlinks=true,allcolors=blue]{hyperref}
\usepackage{physics}
\usepackage[separate-uncertainty=true,exponent-product=\!\cdot\!]{siunitx}

\newcommand{\ntot}{n_\text{tot}}

\newcommand{\funcd}{\boldsymbol{\delta}}
\DeclareMathOperator{\sgn}{sgn}
\DeclareMathOperator{\sh}{sh}
\DeclareMathOperator{\ch}{ch}
\begin{document}

\title{Physical properties of the Hall current}
\author{F. Faisant}
\author{M. Creff}
 \author{J.-E. Wegrowe} \email{jean-eric.wegrowe@polytechnique.edu}
\affiliation{LSI, \'Ecole Polytechnique, CEA/DRF/IRAMIS, CNRS, Institut Polytechnique de Paris, 91120 Palaiseau, France}
\date{\today}

\date{\today}

\begin{abstract}
We study the stationary state of Hall devices composed of a load circuit connected to the lateral edges of a Hall-bar. We follow the approach developed in a previous work (Creff et al. J. Appl. Phys 2020) in which the stationary state of a ideal Hall bar is defined by the minimum power dissipation principle.  The presence of both the lateral circuit and the magnetic field induces the injection of a current: the so-called Hall current. Analytical expressions for the longitudinal and the transverse currents are derived. It is shown that the efficiency of the power injection into the lateral circuit is quadratic in the Hall angle and obeys to the maximum transfer theorem. For usual values of the Hall angle, the main contribution of this power injection provides from the longitudinal current flowing along the edges, instead of the transverse current crossing the Hall bar.
\end{abstract}


\maketitle

\section{Introduction}
\begin{figure} [ht]
   \begin{center}
   \includegraphics[height=9cm]{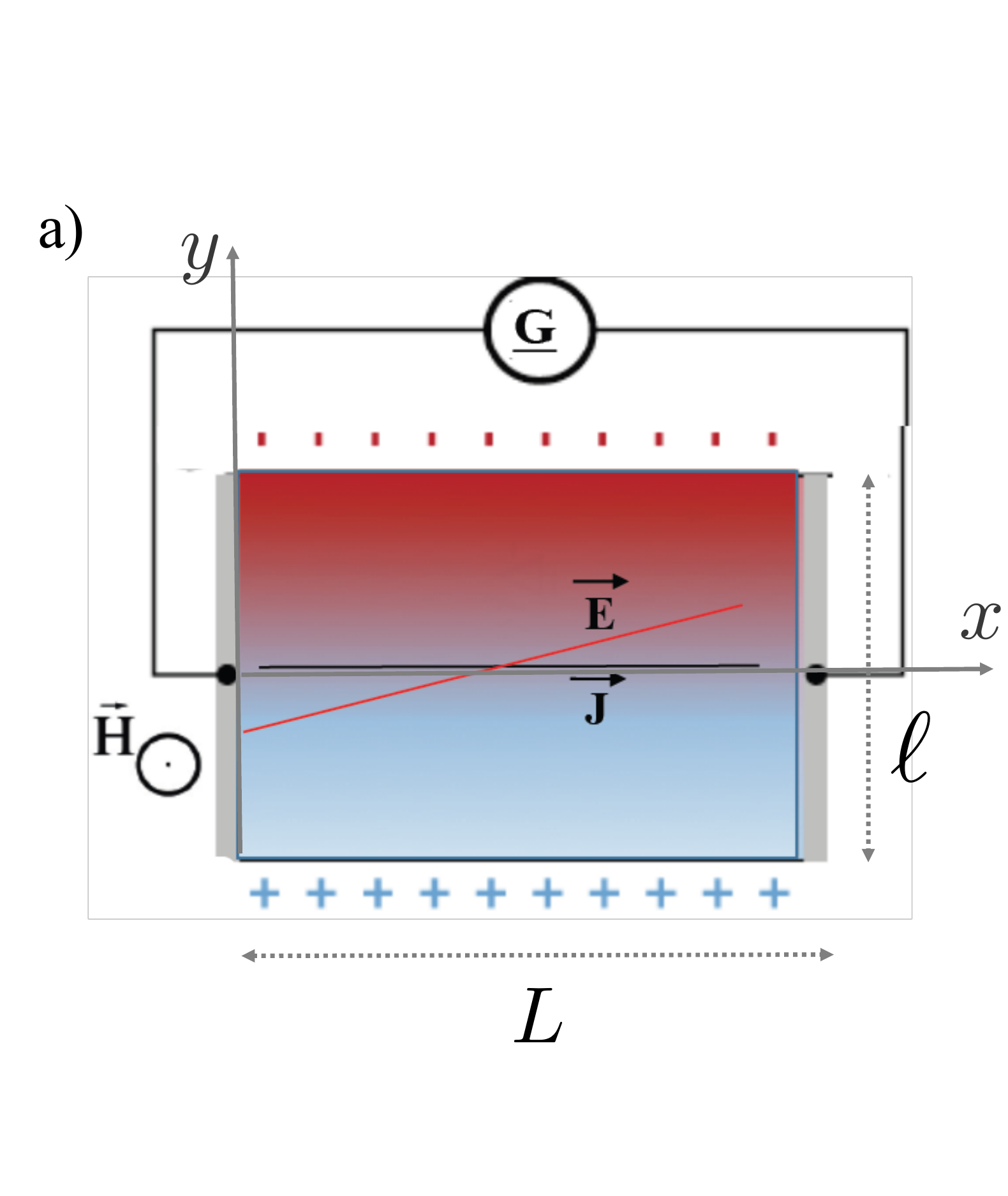}
      \includegraphics[height=8cm]{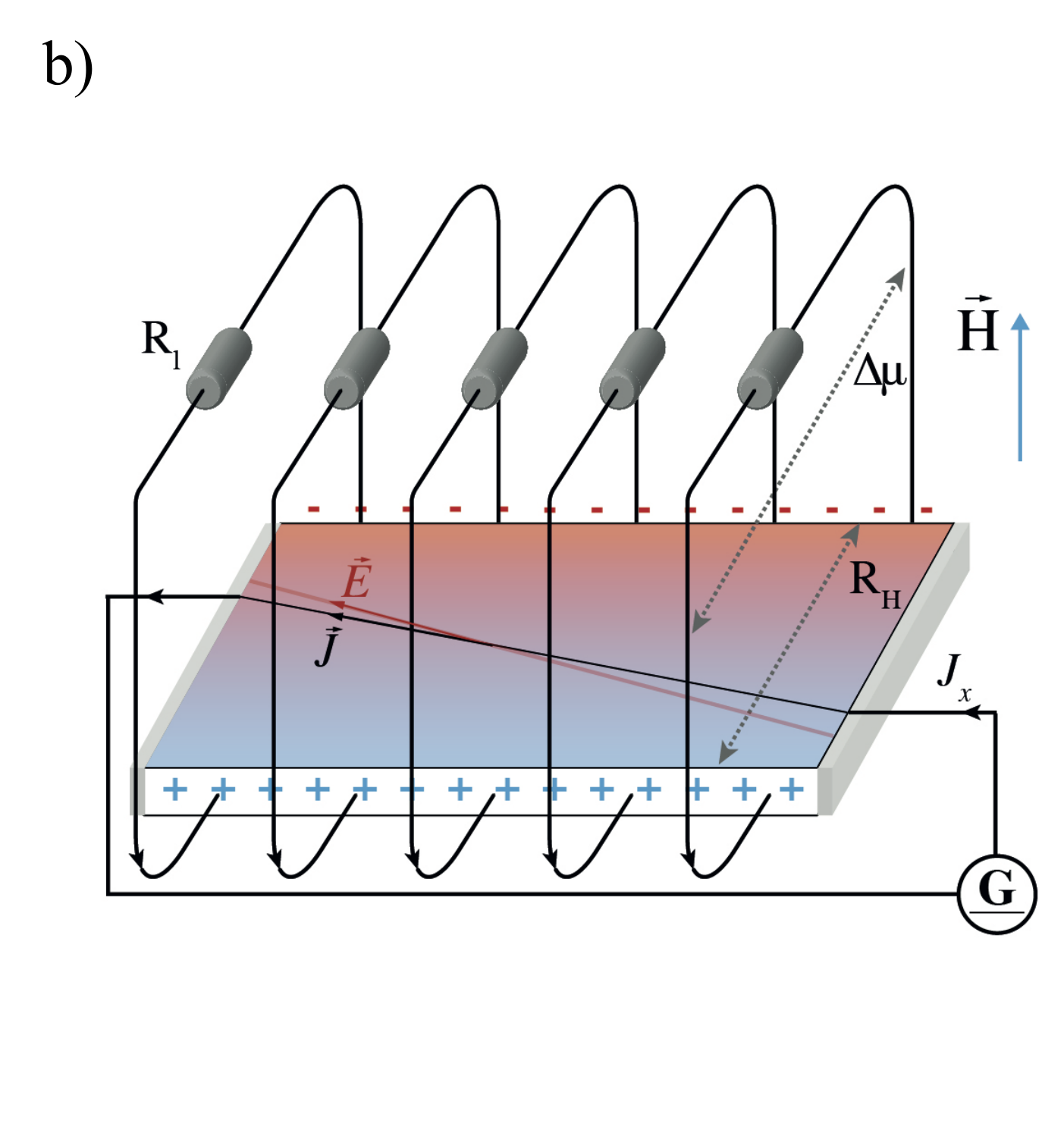}
   \end{center}
   \caption[example] 
   { \label{fig:Fig.1} 
Schematic representation of a Hall-bar with the electrostatic charge accumulation $\pm \delta n$ at the edges, the electric field lines (red) and the current lines (black). The static magnetic field $H$ is applied along the $z$ direction. Note that the scales along $x$ and $y$ are not respected since we need $\ell \ll L$ in order to assume translational invariance along $x$. (a) planar Hall bar without dissipative leakage. (b) Same Hall bar including lateral circuit  with transverse resistance $R_H$ and representation of the load resistance $R_l$ (preserving the translational invariance along $x$). The chemical potential difference $\Delta \mu$ is also represented.}
   \end{figure}

The classical Hall effect \cite{Hall,Corbino} is usually described by the local transport equations for the charge carriers that takes into account the effect of the Laplace-Lorentz force generated by a static magnetic field. Typically, in a planar Hall device, an electric generator imposes a constant electric current $J_x^0$ along the $x$ direction (see Fig.\ref{fig:Fig.1}), and the Hall voltage is then measured transversally along the $y$ direction at stationary regime, as a function of the magnetic field. The physical mechanisms behind this effect and the corresponding transport equations are well-known and are described in all reference textbooks. At stationary state under a perpendicular magnetic field, the {\it  Hall voltage} can be measured, which is due to the accumulation of electric charges between the two edges of the Hall-bar. This state corresponds to a vanishing transverse current - or {\it Hall current} - $J_y = 0$ along the $y$ axis \cite{Aschcroft,Kittel}. Indeed, the accumulation of electric charges at the edges produces a transverse electric field $E_y$ that balances the Lorentz force, so that the system reaches an ``equilibrium'' along the $y$ axis. \\

However, due to the contact with the power generator, the system is not at equilibrium (heat is dissipated), and the presence of the magnetic field is likely to couple the two directions $x$ and $y$ of the device (assumed to be planar), as shown by the transport equations. The reason why - or under what conditions - the system imposes a vanishing Hall current $J_y = 0$ at stationary regime is given by a variational principle: the current distributes itself so as to minimize the Joule heating. A stationary state with $J_y \ne 0$ occurs in some specific situations, that are for instance : (i) the Corbino disk under a magnetic field \cite{Corbino}, (ii) the spin-Hall effect, in which the effective magnetic field is defined by the spin-orbit scattering (presence of a pure spin-current) or (iii) the case of an electric contact that links the two opposite edges to a load resistance. This last situation is present while measuring the Hall voltage, since the internal resistance of a real voltmeter is finite.  \\

The investigation of the condition $J_y = 0$ in a ideal Hall bar was the object of previous publications \cite{Benda,JAP,JAP2}, in which the variational method used in the present work was developed.
Beyond, the case (i) of the Corbino disk is well-known: in the presence of the static magnetic field, an orthoradial current is indeed flowing perpendicular to the radial electric field. The power dissipated in the stationary state is higher than for the equivalent Hall bar \cite{Benda,Madon}. The case (ii) is still controversial  \cite{EPL1,EPL2} and will not be discussed here. 
The question (iii) seems to be disregarded in the literature, but it could be related to the so-called current mode in Hall devices \cite{Popovic}. However, the measured ``Hall current'' is usually an effect of the non-uniform current-lines, due e.g. to misalignment of the metallic electrodes \cite{New1,New2}. In contrast, the goal of this report is to study the physical properties for the configuration that corresponds to the highest symmetry of the device, compatible with the constraints applied to it.

\section{Joule dissipation}

The system under interest is studied in the context of non-equilibrium thermodynamics \cite{Onsager, Onsager_Diss,Bruers,MinDiss,DeGroot,Rubi}. It is a thin homogeneous conducting layer of length $L$ and width $\ell$ contacted to an electric generator, and submitted to a constant magnetic field $\vec H$ oriented along the $z$ axis (see Fig.\ref{fig:Fig.1}). 
 We assume that the conducting layer is planar, invariant by translation along the $x$ axis $\ell \ll L$  (this excludes the region in contact with the power generator), and the two lateral edges are symmetric. 
 
Let us define the distribution of electric charge carriers by $n(y) = n_0 + \delta n(y)$, where $\delta n(y)$ is the charge accumulation and $n_0$ the homogeneous density in the electrically neutral system (e.g. density of carriers without the magnetic field). The charge accumulation is governed by the Poisson's equation $\nabla^2 V = -\frac{q}{\varepsilon} \delta n$, where $V$ is the electrostatic potential, $q$ is the electric charge, and $\varepsilon$ is the electric permittivity. The local electrochemical potential $\mu(x,y)$ - that takes into account not only the electrostatic potential $V$ but also the energy (or the entropy) responsible for the diffusion - is given by the expression \cite{DeGroot,Rubi}  (local equilibrium is assumed everywhere): 
\begin{equation}
  \mu = \frac{k T}{q} \ln \left( \frac{n}{n_0} \right) + V ,
  \label{mu}
\end{equation}
where $k$ is the Boltzmann constant and the temperature $T$ is the temperature of the heat bath in the case of a non-degenerate semiconductor, or the Fermi temperature $T_F$ in the case of a fully degenerate conductor \cite{Rque}. Poisson's equation now reads
\begin{equation}
  \nabla^2 \mu - \lambda_D^2 \frac{q}{\varepsilon} n_0 \nabla^2 \ln \left( \frac{n}{n_0} \right) + \frac{q}{\varepsilon} \delta n = 0,
  \label{poisson-mu}
\end{equation}
where $\lambda_D= \sqrt{\frac{k T \varepsilon}{q^2 n_0}}$ is the Debye-Fermi length. On the other hand, the transport equation under a magnetic field is given by the Ohm's law:
\begin{equation}
\vec{J} = -\hat{\sigma} \vec{\nabla} \mu = - q n \hat{\eta} \vec{\nabla} \mu,
\label{Ohm}
\end{equation}
 where the transport coefficients are the conductivity tensor $\hat{\sigma}$ or the mobility tensor $\hat{\eta}$. In two dimensions and for isotropic material, the mobility tensor is defined by Onsager relations \cite{Onsager}:
\begin{equation*}
  \hat{\eta} =
  \begin{pmatrix}
    \eta    & \eta_H \\
    -\eta_H & \eta
  \end{pmatrix}
  = \eta
  \begin{pmatrix}
    1         & \theta_H \\
    -\theta_H & 1
  \end{pmatrix}
  \quad \text{ with } \quad \theta_H = \frac{\eta_H}{\eta},
\end{equation*}
where $\eta$ is the ohmic mobility, $\eta_H$ the Hall mobility (usually proportional to the magnetic field $\vec{H}=H\vec{e}_z$) and $\theta_H$ the Hall angle. 
The electric current then reads: 

$ \vec{J} = - q n \eta \left( \vec{\nabla} \mu - \theta_H \, \vec{e_z} \times \vec{\nabla} \mu \right)$ (where $\times$ denotes the cross product), or:
\begin{align}
  -q n \eta (1+\theta_H^2) \partial_x \mu & = J_x - \theta_H J_y 
  \label{relations-dxmu} \\
  -q n \eta (1+\theta_H^2) \partial_y \mu & = J_y + \theta_H J_x 
  \label{relations-dymu} \\
\|\vec J\|^2  \equiv J_x^2 + J_y^2 &  = (qn \eta)^2 \, (1 + \theta_H^2) \,  \|\vec \nabla \mu\|^2.
\label{OhmHall3}
\end{align}

The expression of the Joule power dissipated by the system  reads:
\begin{equation*}
  P_J  = S_\text{lat} \int_{-\ell}^{\ell}  q n \eta \|\vec{\nabla} \mu\|^2 \dd{y}
       = \frac{L}{q n_0 \eta (1+\theta_H^2)}\int_{-\ell}^{\ell} \frac{n_0}{n} \|\vec{J}\|^2 \dd{y}.
\end{equation*}
where $S_\text{lat}$ is the lateral surface of the Hall bar (product of the length $L$ by the thickness), and $2 \ell$ is the width.

\section{The ideal Hall bar}

The stationary state is defined by the {\it least dissipation principle}, that states that the current distributes itself so as to minimize Joule heating $P_J$ compatible with the constraints \cite{Onsager_Diss,Bruers,MinDiss}. 

Due to the symmetry of the device and the global charge conservation we have $\int_{-\ell}^{+\ell} \delta n \dd{y} = 0$, and the total charge carrier density is constant $\ntot = \frac{1}{2 \ell}\int n \dd{y}$. For the sake of simplicity, we assume a global charge neutrality so that $\ntot = n_0$. On the other hand, the global current flowing in the $x$ direction throughout the device is also constant along $x$ by definition of the galvanostatic condition. 
The two global constraints read: 
 \begin{equation}
 \int_{- \ell}^{\ell} n(y) \dd{y} = 2 \ell n_0
\quad \textrm{and} \quad 
\int_{- \ell}^{\ell} J_x(y) \dd{y} = 2 \ell J_x^0
\label{Constr}
 \end{equation}
We define for convenience the reduced power $\tilde{P_J} = \frac{q \eta (1+\theta_H^2)}{S_\text{lat}} \, P_J  = \int_{- \ell}^{\ell} \frac{J_x^2 + J_y^2}{n} \dd{y}$.
Let us introduce the two Lagrange multiplayers $\lambda_J$ and  $\lambda_n$ corresponding to the two constraints Eqs(\ref{Constr}). The functional to be minimized then reads: 
\begin{equation}
\tilde {\mathcal{P}}_J [J_x,J_y,n] =   \int_{- \ell}^{\ell} \left ( \frac{J_x^2 + J_y^2}{n}  - \lambda_J \, J_x - \lambda_n \, n \right ) \dd{y}
 \end{equation}
The minimum corresponds to:
 \begin{equation}
  \frac{\funcd \tilde{\mathcal{P}}_J}{\funcd J_x} = 0  \ \Longleftrightarrow \ 2 J_x = n \lambda_J,
 \label{CondJx00}
\end{equation}
 \begin{equation}
  \frac{\funcd \tilde{\mathcal{P}}_J}{\funcd J_y} = 0 \ \Longleftrightarrow \ J_y = 0,
\end{equation}
 \begin{equation}
  \frac{\funcd \tilde{\mathcal{P}}_J}{\funcd (n)} = 0 \ \Longleftrightarrow \ J_x^2 + J_y^2 = - \lambda_n n^2,
 \label{Condn}
\end{equation}

Using Eqs.(\ref{Constr}) and Eq.(\ref{CondJx00}) leads to $\lambda_J = \frac{2 J_x^0}{n_0}$ so that $J_x  = \frac{n}{n_0} J_x^0 $ (and from Eq.(\ref{Condn}) we have furthermore : $\lambda_n = - (J_x^0 / n_0)^2$). Hence, the minimum is reached for 
\begin{equation}
  J_x(y) = J_x^0\frac{n(y)}{n_0} \quad \text{and} \quad J_y = 0,
  \label{min}
\end{equation}

The usual stationarity condition $\vec \nabla \cdot \vec J = 0$ is verified. Inserting the solution (\ref{min}) into the transport equations (\ref{relations-dxmu},\ref{relations-dymu}), we deduce $\partial_x \mu = \frac{-J_x^0}{q n_0 \eta (1+\theta_H^2)}$ and $\partial_y \mu = \frac{\theta_H J_x^0}{q n_0 \eta (1+\theta_H^2)}$. These two terms are constant so that the electochemical potential of the stationary state is harmonic: $\nabla^2 \mu = 0$. Since the profile of the lateral current $J_y(y)$ is defined by the charge density $n(y)$, the Poisson's equation Eq.(\ref{poisson-mu}) for $\nabla^2 \mu = 0$ gives the solution:
\begin{equation}
  \lambda_D^2 \partial_y^2 \ln \left(1 + \frac{\delta n }{n_0} \right) = \frac{\delta n}{n_0}.
  \label{poisson-final}
\end{equation}
Once again, the boundary conditions for the density $n$ are not defined locally but globally by Eq.({\ref{Constr}}), and by the integration of the Gauss's law $\vec{\nabla} \cdot \vec{E} = \partial_y E_y = \frac{q}{\varepsilon} \delta n$, at a point $y_0$ (see Appendix C in reference \cite{JAP}):
\begin{equation}
  E_y(y_0) = - \partial_y V(y_0) = - \frac{q}{2 \varepsilon} \int_{-\ell}^{\ell} \delta n(y) \sgn(y-y_0) \dd{y} + \Delta E^{\infty},
  \label{ElecField}
\end{equation}
where the constant $\Delta E^{\infty} = E(+ \infty) + E(- \infty)$ accounts for the electromagnetic environment of the Hall device ($\Delta E^{\infty} = 0$ in vacuum) and the Sign function $\sgn(y-y_0) \equiv (y-y_0)/\abs{y-y_0}$ accounts for the opposite sign of the charge accumulation at both edges. Inserting the stationary solution (\ref{min}) and the relation (\ref{relations-dymu}) for $\partial_y\mu$ gives the condition:
\begin{equation}
  \frac{2 \theta_H J_x^0 C_0}{1 + \theta_H^2} + 2 \lambda_D^2 \partial_y \ln(\frac{n}{n_0})(y_0) +2 C_E +  \int_{-\ell}^{\ell} \delta n(y) \sgn(y-y_0) \dd{y} = 0,
  \label{condition}
\end{equation}
where $C_0 = \frac{\varepsilon }{q^2 n_0 \eta}$ and $C_E = \frac{\varepsilon \Delta E^{\infty}}{q n_{0}}$. Using this condition and fixing $n_{0}$ gives a unique solution for $n(y)$, and the stationary current Eq.(\ref{min}) is fully determined.

This derivation was the object of the report published in reference \cite{JAP}, and the result was confirmed by an independent stochastic approach\cite{JAP2}. For small Debye length $\lambda_D/\ell \ll 1$, the charge accumulation $\pm \delta n$ at the edges give rise to the voltage $V_H^0= \frac{\theta_H 2 \ell J_x^0}{q n_0 \eta}$.
For low magnetic field $H$ we have $\theta_H \approx \eta H$ and the usual expression of the Hall voltage is recovered:  $V_H^0= \frac{H 2 \ell J_x^0}{q n_0}$.

\section{Effect of a lateral passive circuit}
\label{sec:sections}

The solution found in the preceding section is valid as long as the dissipation due to charge leakage at the edges is negligible with respect to the dissipation inside the device. However, if it is no longer the case, the stationary regime should be reconsidered by introducing the dissipation due to the resistance of a lateral passive circuit that connects the edges of the Hall bar.
In order to take into account this supplementary dissipation, we introduce the load conductivity $g$ ($\Omega^{- 1} \cdot \mathrm{m}^{- 2}$) of the lateral circuit (see Fig.\ref{fig:Fig.1}b). The power dissipated in the lateral circuit is, by definition of $g$:
\begin{equation*}
  P_\text{lat}  \ = \  S_\text{lat}\, g\, \Delta \mu^2
  \label{Power_lat}
\end{equation*}
where $\Delta \mu = \mu (+ \ell) - \mu (- \ell)$ is the difference of the chemical potential between both edges (see Fig.\ref{fig:Fig.1}b). We assume that the load conductivity $g$ does not depend on the magnetic field. From a topological point of view, despite the presence of electric charge accumulation at the edges $\delta n \ne 0$, the fact that the system is doubly connected - instead of simply connected - suggests that the corresponding device is closer to a Corbino disk than a Hall bar \cite{Benda}. 
 \\
Note that due to our hypothesis of the invariance along $x$, we do not treat the case of a unique wire that joints the two edges of the Hall bar, that would form two ``punctual'' contacts on both edges (see Fig.\ref{fig:Fig.1}b). Indeed, such a contact would break the translation invariance symmetry along $x$, and would distort the current lines in a specific manner that depends on the details of the contact geometry and resistivity. Such a contact-specific effect is not related the generic problem studied here. Incidentally, it is well-known that the main advantage of the Corbino disk with respect to the Hall-bar device is precisely that it is much easier to design two quasi-perfect concentric equipotentials (circular symmetry) instead of two quasi-perfect longitudinal equipotentials (translational symmetry). 

Using Eq.(\ref{relations-dymu}), the difference of chemical potential can be expressed as a function of the current:
\begin{equation}
\Delta \mu \  = 
   \  \int_{- \ell}^{+ \ell} \dd{y}\, \partial_y
   \mu \  =\  \int_{- \ell}^{+ \ell} \dd{y}\,
   \frac{J_y + \theta_H J_x}{qn \eta (1 + \theta_H^2)}
\end{equation}
so that
\begin{equation}
 P_\text{lat} \  = \  \frac{S_\text{lat}\,g}{(q \eta)^2  (1 + \theta_H^2)^2}  \left( \int_{-
   \ell}^{+ \ell} \dd y \frac{J_y + \theta_H J_x}{n} \right)^2
   \label{PowerLeak}
\end{equation}

As in the preceding section, we define the reduced power $\tilde{P} = \frac{q \eta (1 + \theta_H^2)}{S_\text{lat}} P$. The total power dissipated is then:
\begin{equation}
    \tilde{P} \  = \  \tilde{P}_J + \tilde{P}_\text{lat} \  = \  \int_{- \ell}^{+ \ell} \dd
    y \frac{J_x^2 + J_y^2}{n} \  +
    \  \alpha  \left( \frac{n_0}{2 \ell} \int_{- \ell}^{+ \ell} 
    \dd y \frac{J_y + \theta_H J_x}{n} \right)^2
    \label{Power}
\end{equation}
where we have introduced the {\bf dimensionless control parameter} $\alpha$:
\begin{equation}
\alpha = \frac{2 \ell\, g } {q \eta n_0 (1 + \theta_H^2)}
\label{alpha}
\end{equation}
Note that the control parameter $\alpha$ is the ratio $\alpha = \frac{R_H}{R_l}$ of the ``Hall resistance'' per surface unit $R_H \equiv \frac{V_H}{J_x^0} = \frac{2 \ell}{q n_0 \eta(1 + \theta_H^2)}$ over the resistivity $R_l = \frac{1}{ g}$ of the load.
 
Accordingly, the minimization of the corresponding functional $\tilde{\mathcal{P}}$ now reads:
\begin{equation}
  \frac{\funcd \tilde{\mathcal{P}}}{\funcd J_x} = 0 \ \Longleftrightarrow \
 2 \,  \alpha \, A \, \theta_H + 2  J_x  = n \lambda_J,
 \label{CondJx0}
\end{equation}
where we have defined for convenience the constant $A \equiv \frac{n_0}{2 \ell} \int_{-\ell}^{\ell} \frac{J_y + \theta_H J_x}{n} \dd{y}$. Furthermore:
 \begin{equation}
  \frac{\funcd \tilde{\mathcal{P}}}{\funcd J_y} = 0 \ \Longleftrightarrow \
\alpha \, A + J_y   = 0,
 \label{Jy0}
\end{equation}
and
 \begin{equation}
  \frac{\funcd \tilde \tilde{\mathcal{P}}}{\funcd (n)} = 0 \ \Longleftrightarrow \
  2 \alpha \, A + J_x^2 + J_y^2 = - \lambda_n \, n^2,
 \label{CondnB}
\end{equation}
Equations (\ref{CondJx0}) and (\ref{CondnB}) define the Lagrange multipliers $\lambda_J$ and $\lambda_n $ and will not be used in the following. From Eq.(\ref{Jy0}) we can immediately deduce that :

\begin{itemize}

\item
$J_y$ does not depend on $y$. 

\item
In the absence of a magnetic field, $\theta_H = 0$, and we have $\frac{n_0}{2 \ell} \alpha \, J_y  \int  \tfrac{\dd y}{n} + J_y = 0$, and $J_y = 0$ is the unique solution (since $\alpha$ and $n$ are positive).

\item
If the load resistance goes to infinity $R_l \to \infty$ (or $g \to 0$), the power dissipated by the current leakage is negligible and we are back to the case discussed in the preceding section: the stationary state is defined by $ J_x(y) = J_x^0\frac{n(y)}{n_0} $ and $J_y = 0$.

\item
In the case of a short-circuit by the edges (i.e. the case of a Corbino disk), $R_l \to 0$ (or $g \to \infty$), we have $ A \equiv \int \frac{J_y + \theta_H J_x}{n} \dd{y} \to 0$, which leads to the solution, at the limit:
$J_y = - \theta_H J_x $. This is indeed the well-known stationary state for the Corbino disk, which corresponds to the maximum current $J_y$ \cite{Benda}.

\end{itemize}

\section{Between Corbino disk and Hall bar}

\begin{figure} [ht]
   \begin{center}
   \includegraphics[height=8cm]{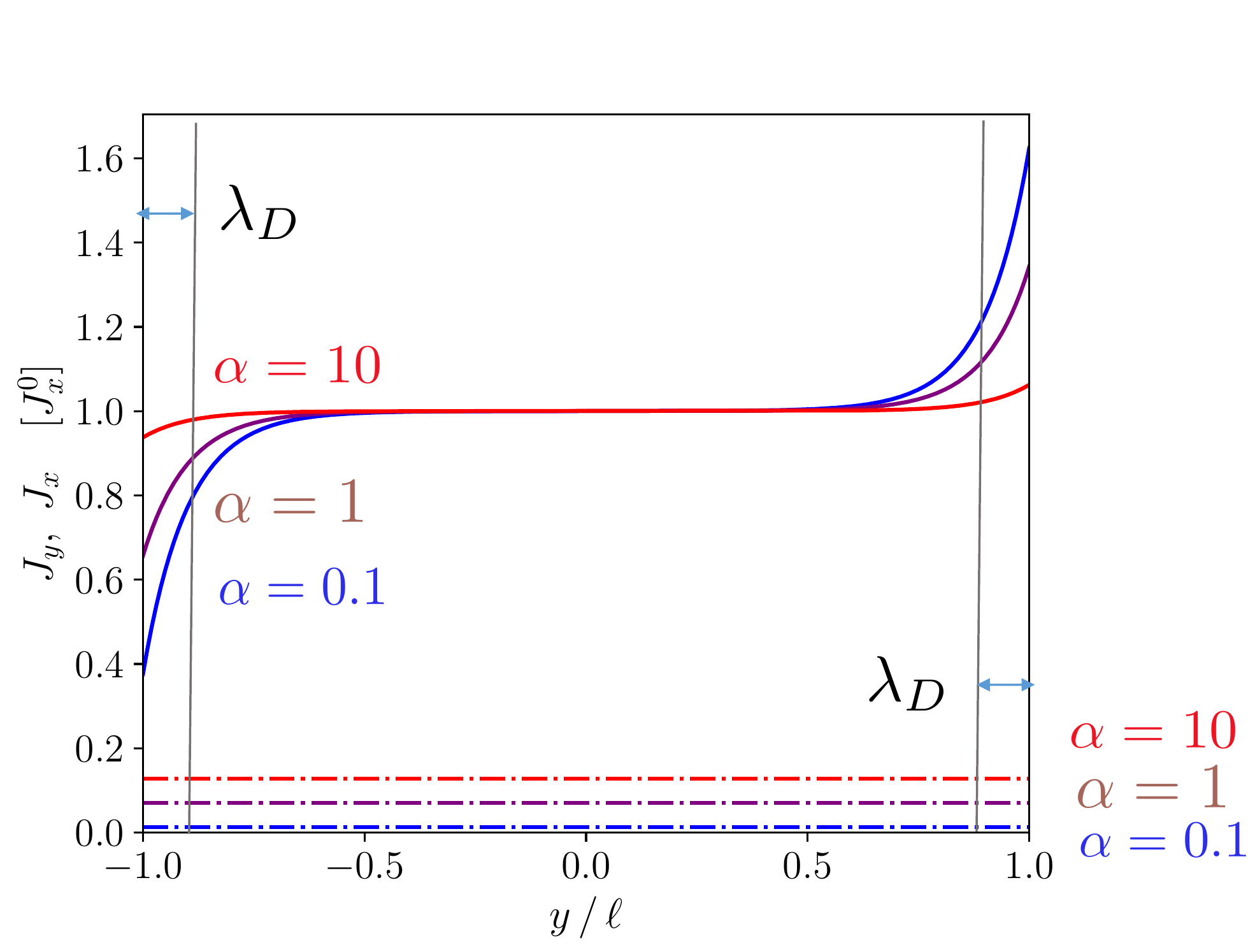}
   \end{center}
   \caption[example] 
   { \label{fig:Fig.2} Typical profiles for the longitudinal current $J_x(y)$ (plain lines) and homogeneous transversal current $J_y$ (dotted horizontal lines) across the Hall-bar for various values of the parameter $\alpha = \{0.1, 1,10\}$, and in units of the injected current $J_x^0$. The Debye length  $\lambda_D$ is indicated by the vertical lines.} 
   \end{figure}

Introducing the constant current inside the integral of Eq.(\ref{Jy0}) with $J_y \equiv \int_{- \ell}^{+ \ell}
  \frac{J_y \dd y}{2 \ell}$, and dividing by $\frac{J_y}{2 \ell}$ (for $J_y \ne 0$), we obtain 
 \begin{equation}
    \int_{- \ell}^{+ \ell} \dd y \left( 1 + \alpha \frac{n_0}{n}  \left( 1
    + \tfrac{\theta_H}{J_y} J_x \right) \right) \  = \  0
    \label{JyJx}
  \end{equation}
As pointed-out above, the two limiting cases are solution of Eq.(\ref{JyJx}). At the limit of the perfect Hall bar (defined by an infinite load resistance and $\alpha = 0$), a vanishing transverse current $J_y \rightarrow 0$ is recovered, while at the limit of the perfect Corbino disk (defined by $R_l =0$ or $\alpha = \infty$) the Corbino current $J_y = - \theta_H J_x $ is recovered. Without loss of generality, the solution $J_y(\alpha)$ can be expressed with introducing an arbitrary function $f(\alpha)$ such that $J_y =  - f(\alpha) \, \theta_H J_x^0$. The function $f(\alpha)$ can be determined by using the {\it sufficient} condition
\begin{equation}
1 + \alpha \frac{n_0}{n}  \left( 1+ \theta_H \frac{J_x}{J_y} \right) = 0.
\label{local-condition-Jx-Jy}
\end{equation}
We then obtain $J_x(y) = J_x^0\, f(\alpha) \left (1 + \frac{1}{\alpha} \frac{n(y)}{n_0} \right )$. Applying the two global contraints Eqs.(\ref{Constr}) leads to the expression $f(\alpha) = \frac{\alpha}{\alpha +1} \in \, [0,1]$, and thus to:
\begin{equation}
J_y = - \frac{\alpha}{\alpha + 1 } \theta_H J_x^0.
\label{Jyleak}
\end{equation}
which interpolates the two limiting regimes for arbitrary ratio $\alpha = R_H/R_{\ell}$. From Eq.(\ref{local-condition-Jx-Jy}) we deduce:
 \begin{equation}
J_x (y) = \frac{J_x^0}{\alpha + 1 } \left (\alpha + \frac{n(y)}{n_0}  \right ).
\label{Jxleak}
\end{equation}
The lateral current $J_y$ is homogeneous (it does not depend on $y$), while the corresponding longitudinal current $J_x(y)$ is non-uniform and follows the profile of the charge accumulation $n(y)$. The relation $\vec{\nabla} \cdot \vec{J} = 0$ is still verified, but the chemical potential is no longer harmonic. The derivative $\partial_y \mu = -\frac{J_x^0}{\alpha+ 1} \, \frac{\theta_H}{q \eta  n_0 (1+\theta_H^2)}$ is still constant (decreased by a factor $1/(\alpha +1)$) while $\partial_x \mu(y)$ now depends on $y$.
The typical profiles of the longitudinal and transverse currents Eq.(\ref{Jxleak}) and Eq.(\ref{Jyleak}) are plotted in Fig.\ref{fig:Fig.2} in unit of the injected current $J_x^0$. 
 
The Hall voltage with lateral load resistance can be derived easily. Inserting the solution Eqs.(\ref{Jyleak},\ref{Jxleak}) instead of Eqs.(\ref{min}) into Eq.(\ref{condition}), only the expression of the parameter $C_0$ is modified by the factor $1/( 1 + \alpha)$:
\begin{equation}
C_0(\alpha) = \frac{\epsilon}{q^2 n_0 \eta} \, \frac{1}{1 + \alpha}
\label{C0alpha}
\end{equation}

Assuming $C_E=0$, the charges accumulation $\delta n /n_0$ at the edges is reduced by the same factor: 
\begin{equation}
\frac{\delta n(\alpha)}{n_0} =    \, \frac{\mathcal C}{1 + \alpha}
\label{charge_acc}
\end{equation}
where $\mathcal C = \frac{\delta n}{n_0}(\alpha = 0)$ is the charge accumulation without lateral circuit, as calculated in reference \cite{JAP}: 

$\mathcal C \equiv \frac{\theta_H J_x^0 C_0}{n_0(1 + \theta_H^2)} \frac{1}{\lambda_D} \frac{\sh(y/\lambda_D)}{\ch(y/(2\lambda_D))}$.
For a vanishing screening length $\lambda_D \rightarrow 0$, the charge accumulation reduces to Dirac distributions at the edges of the Hall bar \cite{JAP}:
\begin{equation}
  q\, \delta n(y) = \sigma^\mathsf{S} \, \big( \delta \left (y - \ell \right) - \delta \left (y + \ell \right ) \big)
  \label{charges}
\end{equation}
where $ \sigma^\mathsf{S}$ is the surface charge:
\begin{equation}
  \sigma^\mathsf{S}(\alpha) = J_x^0 \, \frac{ \theta_H}{1+\theta_H^2}  \frac{\varepsilon }{q n_0 \eta} \frac{1}{1 + \alpha},
   \label{conductivity}
\end{equation}
 that does not depend on $J_y$.
Assuming the usual low magnetic field limit, we have $\theta_H \approx \eta H$ and the Hall voltage is deduced:
\begin{equation}
  V_H(\alpha) = \frac{\sigma^\mathsf{S} L}{\varepsilon} =  \frac{J_x^0 H L}{q n_0} \frac{1}{1 + \alpha}
  \label{Hallvoltage}
\end{equation}
 
The voltage Eq.(\ref{Hallvoltage}) divided by Hall voltage $V_H^0$ of the ideal Hall bar is simply given by $\frac{V_H}{V_H^0} = \frac{1}{1 + \frac{R_H}{R_l}}$, where we have replaced the parameter $\alpha$ by its value $\alpha = R_H/R_l$.\\

\begin{figure} [ht]
   \begin{center}
   \includegraphics[height=8cm]{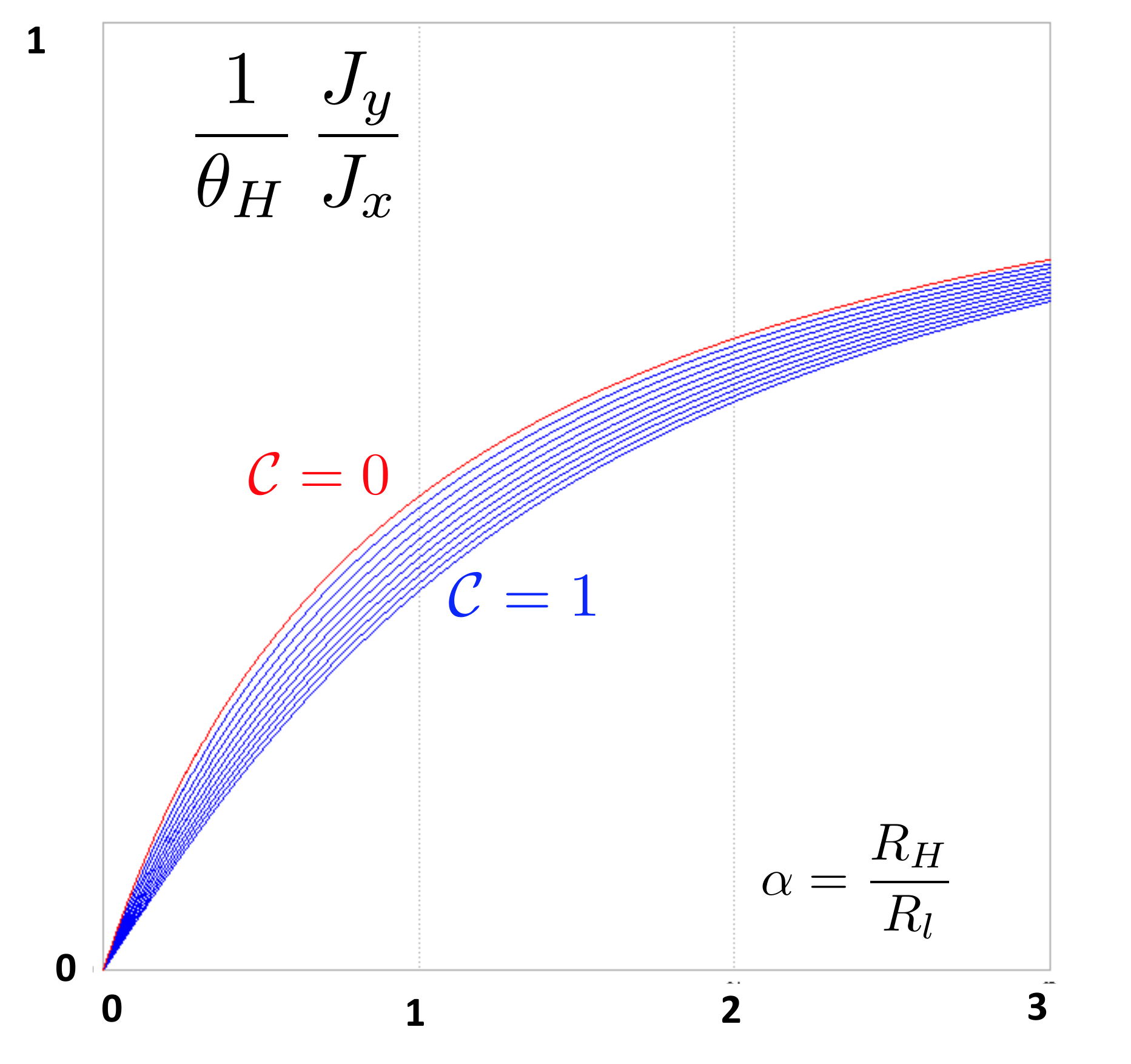}
   \end{center}
   \caption[example] 
   { \label{fig:Fig.3} The ratio of the transverse current over longitudinal current divided by $\theta_H$: $J_y(\alpha)/ \left (\theta_H  \,  J_x(\alpha) \right)$ at the edge ($y = \ell$) is plotted as a function of the parameter $\alpha =R_H/R_l$ for different values of the initial charge accumulation $\mathcal C = \delta n / n_0(\alpha = 0)$ (varying from $0$ to $1$ with step $0.1$).}
   \end{figure} 

 Note that the ratio if the transverse current over longitudinal current:
\begin{equation}
\frac{J_y}{J_x}(\alpha) = \frac{ \theta_H \, \alpha }{ \alpha + 1 + \frac{ \mathcal C}{1 + \alpha}}
\label{RatioJyJx}
\end{equation}
is small for usual values of the angle $\theta_H$. This ratio divided by $\theta_H$ is plotted in Fig.\ref{fig:Fig.3} at the edge $y = \ell$, as a function of $\alpha = R_H/R_l$. The quantitative study of the result Eq.(\ref{RatioJyJx}) shows that the power injected into the lateral circuit is mainly carried by the longitudinal current $J_x(y)$ instead of the transverse current $J_y$. Indeed, as shown by Eq(\ref{charges}) and (\ref{conductivity}), the system can be interpreted as a capacitor which is recharged permanently by the longitudinal current $J_x$ only, in order to keep the charge accumulation $\delta n$ at stationary state. In other terms, the electric charges that are injected into the external circuit are mainly due to the discharge of the lateral edges, resupplied permanently by the longitudinal current $J_x$. This rather counter-intuitive picture invalidates that of a Hall current $J_y$ composed of carriers of charge carriers flowing transversally from one edge to the other through the Hall bar.\\

However, if the parameter $\alpha$ is large enough, the contribution of the transverse current $J_y$ to the total current becomes sizable for small values of the load resistance $R_l \ll R_H$ in nearly intrinsic semiconductors. Typically, the value of $\theta_H \approx 0.14$ is obtained in a field of $\SI{1}{\tesla}$ in Silicon with impurity density of about $10^{15}$ $\mathrm{cm}^{-3}$. The transverse current $J_y$ injected into the lateral circuit can then reach the amplitude of the longitudinal current $J_x$ for a magnetic field of the order of $\SI{1}{\tesla}$ if the coefficient $\alpha$ is of the order of $100$. The load circuit is then {\it close to a short-circuit} between the two edges of the Hall-bar, and the corresponding device is like a Corbino disk, i.e. a device in which the charge accumulation is not allowed.

\section{Power injected}
The total power $\tilde P = \tilde P_J + \tilde P_\text{lat}$ - given in Eq.(\ref{Power}) - is the sum of the Joule heating $\tilde P_J$ dissipated inside the Hall device, and the power $\tilde P_\text{lat}$ dissipated into the lateral passive circuit. Inserting the stationary state Eqs.(\ref{Jyleak},\ref{Jxleak}) and using the first global condition in Eqs.(\ref{Constr}), we obtain:
 \begin{equation}
\tilde P(\alpha) = \frac{\left ( J_x^0\right )^2}{(\alpha + 1)^2} \left (\alpha^2 (1 + \theta_H^2) \int_{-\ell}^{+ \ell} \frac{\dd{y}}{n} + (2 \alpha + 1) \frac{2 \ell}{n_0}  + \frac{2 \ell \alpha}{n_0} \theta_H^2 \right )
\label{Power_st}
 \end{equation}
  
Assuming that $\delta n \ll n_0$, we have $\int_{-\ell}^{+ \ell} \frac{\dd{y}}{n(y)} \simeq 2 \ell/n_0$ and the total dissipated power reads:
\begin{equation}
P_\text{tot}(\alpha) \, \simeq \, \frac{2 \ell \left ( J_x^0 \right )^2 }{q \eta n_0}  \, \frac{\alpha^2 (1 + \theta_H^2) +  \alpha (2 +\theta_H^2) +1}{(\alpha + 1)^2} \, = \, P(0) \, \left (1 + \theta_H^2 \frac{ \alpha}{\alpha + 1}  \right ) 
 \label{Power_Approx1}
\end{equation}
where $P(0)$ is the power dissipated by the ideal Hall-bar without lateral contact.

On the other hand, the power injected into the lateral circuit is 
\begin{equation}
P_\text{lat}(\alpha) = P(0) \, \left (\theta_H^2 \frac{ \alpha}{(\alpha + 1)^2}  \right ) 
 \label{Power_Approx2}
\end{equation}

\begin{figure} [ht]
   \begin{center}
    \includegraphics[height=6cm]{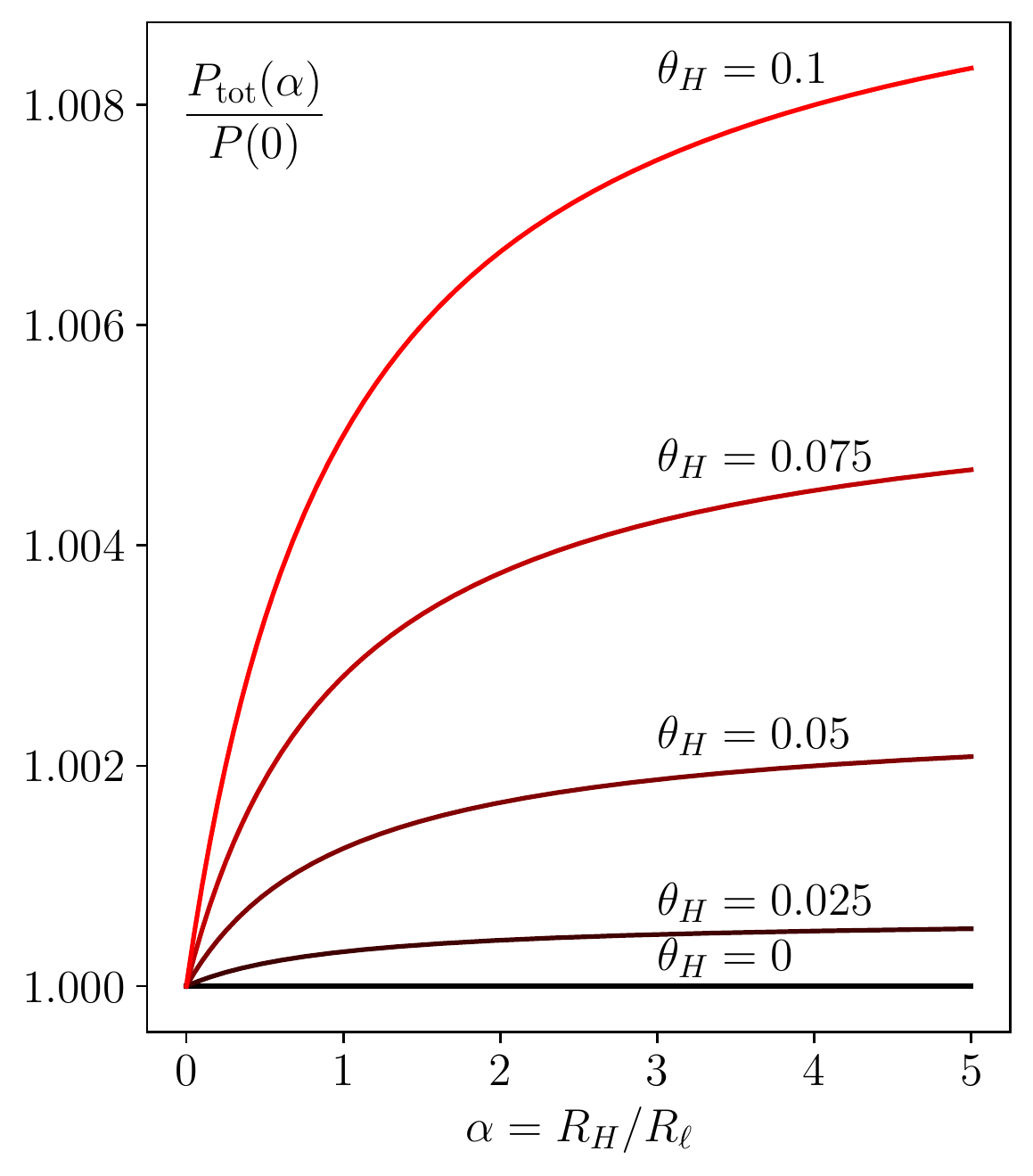} \raisebox{8cm}{(a)} \hspace{5em} 
    \includegraphics[height=6cm]{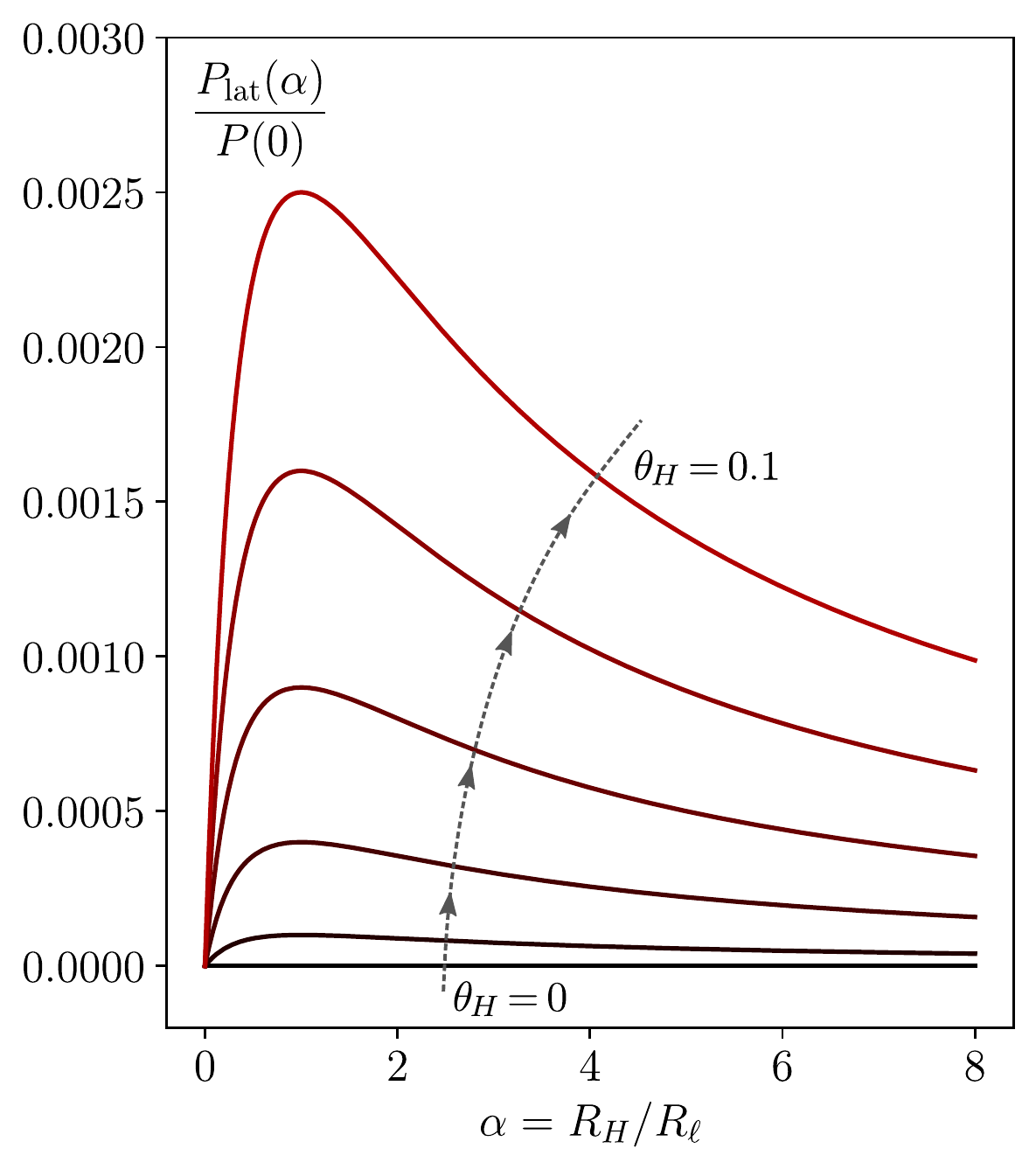} \raisebox{8cm}{(b)} 
   \end{center}
   \caption[example] 
   { \label{fig:power-dissip} 
Amplification of the power as a function of $\alpha =R_H/R_l$ for different values of the Hall angle $\theta_H$ (i.e.\ of the static magnetic field $H$) from $0$ to $0.1$. (a) Total power $P_\text{tot}$ dissipated in the device normalized by the power without lateral circuit $P(0)$. (b) Power $P_\text{lat}$ injected in the lateral circuit normalized by the power without lateral circuit $P(0)$. The maximum coincides with the resistance matching $R_{\ell}=R_H$.}
   \end{figure}

The total power dissipated in the lateral circuit Eq.(\ref{Power_Approx1}) normalized by $ P(0)$ is plotted in Fig.\ref{fig:power-dissip}(a) and the power injected into the lateral circuit Eq.(\ref{Power_Approx2}) normalized by $ P(0)$  is plotted in Fig.\ref{fig:power-dissip}(b), as a function of $\alpha =R_H/R_l$. The different profiles corresponds to different values of $\theta_H$ from $0$ to $0.1$. 
Due to the small values of the Hall angle $\theta_H$, the power injected into the lateral circuit is a small fraction of the total power dissipated by the device. The ratio $P_\text{lat}/P_\text{tot}$ - i.e. the efficiency of the injection - is indeed proportional to $\theta_H^2$.

Note that in Fig.\ref{fig:power-dissip}(b) the power injected into the lateral circuit reaches a maximum at $\alpha=1$, i.e. for $R_\ell = R_H$, independently of the magnetic field. Indeed, the situation is analogous to a voltage source with internal resistance $R_H$, loaded with $R_\ell$. The expression $P_\text{lat}(R_\ell)$ is then an illustration of the so called {\it maximum power transfer theorem}, where the maximal injected power is achieved at the {\it impedance matching condition} $R_\ell = R_H$. This observation gives an intuitive meaning of the Hall resistance $R_H$ as the internal resistance of a voltage source, when the Hall bar is used as the power supply for a lateral circuit.

\section{CONCLUSION}

We have performed a quantitative analysis of the stationary state of a Hall-bar connected to a load circuit at the lateral edges. This configuration corresponds to the so-called {\it current mode} of Hall devices. This analysis is based on a variational approach developed in previous works. The model assumes a planar device, a perfect symmetry of the two lateral edges, and a translational invariance along the longitudinal direction $x$ (the deformation of the current lines due to the contacts is not taken into account). The expression of the non-uniform longitudinal current $J_x(y)$ is calculated. This current allows the charge accumulation to be maintained at stationary state. When a lateral circuit is connected to the lateral edges of the Hall-bar, it is shown that the current $J_x(y)$ is amplified and a Hall current is generated: $J_y \ne 0$. The power injected from the Hall bar to the lateral circuit can be controlled by the magnetic field and by the load resistance $R_{\ell}$. It is shown that the physical significance of the Hall resistance $R_H$ is that of the usual internal resistance of a voltage source, when the Hall bar is used as the power supply for the lateral circuit.\\

Beyond, the surprising result of this study is that, for usual values of the Hall angle, the main contribution of the power injected into the lateral circuit is due to the longitudinal current $J_x$ instead of the transverse current $J_y$. This means that the device can be interpreted as a capacitor which is recharged permanently by the longitudinal current $J_x$ only, in order to keep the charge accumulation $\delta n$ at stationary state. In other terms, the electric charges that are injected into the external circuit are mainly due to the discharge of the lateral edges, resupplied permanently by the longitudinal current $J_x$. This rather counter-intuitive picture invalidates that of a Hall current  $J_y$ composed of charge carriers flowing transversally from one edge to the other through the Hall bar.
However, this more intuitive Hall-current regime with sizable $J_y$ is able to take place for nearly intrinsic semiconductors (for which $\theta_H \approx 0.15$ or above), for small enough load resistance $R_{ell} < (R_H/100)$: the device is then close to a Corbino disk. The two different regimes are then able to take place in the same device, depending on the values of the load resistance $R_{\ell}$.


\begin{thebibliography}{0}
\bibitem{Rque} The expression is valid for non-degenerate semiconductor (Maxwellian distribution). However, in the case of degenerate metal, the expression is an approximation for $\delta n/n_0  \ll 1$.
\end{thebibliography}
\end{document}